\documentclass[aps,prl,twocolumn,groupedaddress,showpacs,superscriptaddress]{revtex4}

\usepackage{graphicx}
\usepackage{latexsym}
\def\RT3{$R$Te$_3$}
\def\A{$\rm \AA$}
\begin{document}

\title{STM Studies of  TbTe$_3$: Evidence for a fully Incommensurate Charge Density Wave }

\author{A.~Fang}
\affiliation{Department of Applied Physics, Stanford University, Stanford, CA 94305}
\author{N.~Ru}
\affiliation{Department of Applied Physics, Stanford University, Stanford, CA 94305}
\author{ I.~R.~Fisher} 
\affiliation{Department of Applied Physics, Stanford University, Stanford, CA 94305}
\author{A.~Kapitulnik}
\affiliation{Department of Applied Physics, Stanford University, Stanford, CA 94305} 
\affiliation{Department of Physics, Stanford University, Stanford, CA 94305}

\date{\today}

\begin{abstract} We observe unidirectional charge density wave ordering on the cleaved surface of TbTe$_3$ with a Scanning Tunneling Microscope at $\sim$6 K.  The modulation wave-vector $q_{CDW}$ as determined by Fourier analysis is 0.71 $\pm$ 0.02 $\times2\pi/c$.  (Where $c$ is one edge of the in-plane 3D unit cell.) Images at different tip-sample voltages show the unit cell doubling effects of dimerization and the layer below.  Our results agree with bulk X-ray measurements, with the addition of ~(1/3) $\times2\pi/a$ ordering 
perpendicular to the CDW.  Our analysis indicates that the CDW is incommensurate. \end{abstract}

\maketitle

Charge density waves (CDW) in weakly interacting systems 
are predominantly driven by Fermi surface (FS) nesting, where a single vector $q_{CDW}=2k_f$ at the peak of the generalized susceptibility connects points on the FS. In 2D systems, this nesting is enhanced if the FS has parallel regions.  The rare-earth tri-telluride series \RT3 ($R$ = rare-earth) is such a quasi-2D system, formed in the orthorhombic space group $Cmcm$\cite{norling}, and contains double layers of nominally square Te planes separated by $R$Te block layers.  A unidirectional CDW \cite{unidirectional} was first detected in this system by transmission electron microscopy (TEM) \cite{dimasi_1995}. Further measurements of angle resolved photoemission spectroscopy (ARPES) have shown that large portions of the FS nested by 
$q_{CDW}$ are indeed gapped, indicating that the CDW is driven by FS nesting \cite{brouet,gweon}. 

The band structure near the Fermi surface can be modeled using the Te $p_x$ and $p_y$ in-plane bonding orbitals in a single Te plane, which is doped from the block layer.  Such a model has tetragonal $C_4$ symmetry and a Brillouin zone, hereafter referred to as the 2DBZ.  This model fits the measured bands well \cite{brouet} and theoretical investigations \cite{yao} show a nesting along the diagonal of the 2DBZ, with $q_{CDW}\approx(0.75,0.75)\times\pi/a_0$, where $a_0$ is the Te-Te spacing of 3.1 \A.  Yao {\it et al.} \cite{yao} argued that this model has a hidden 1D nature, resulting in a  unidirectional CDW. 
The structural modulation arising from the CDW in  \RT3 has also been observed by X-Ray Diffraction (XRD) \cite{malliakas, ru-transition, malliakas_2006}. Though the CDW exists in the Te planes, the adjacent block layer introduces another super-lattice with in-plane basis vectors ({\bf a}, {\bf c}) that are 45$^o$ to the Te net and $\sqrt{2}$ times longer than $a_0$.\cite{baxis} The new reduced three-dimensional Brillouin zone (3DBZ) has half the in-plane area and is rotated 45$^o$ from the 2DBZ. This additional periodicity  folds (a small amount of) spectral weight from the bands back into the 3DBZ.   For diffraction measurements such as XRD, the wave-vector is ambiguous within a reciprocal lattice vector, and thus is often reported inside the reduced 3DBZ as $q_{CDW}\approx 2/7\times2\pi/c$.  On the other hand, ARPES reports $q_{CDW}\approx 5/7\times2\pi/c$ based on gaps in the FS. \cite{brouet}. Thus there exists some ambiguity regarding which modulation is the dominant contribution to the CDW.
Regarding the structure of the CDW, Kim {\it et al.} in a recent paper employed atomic pair distribution function (PDF) analysis of powder X-Ray data supplemented with local STM measurements and suggested that the modulation in CeTe$_3$ is composed of commensurate domains separated by discommensurations.\cite{michigan} 
 
\begin{figure}[h]
\includegraphics[width=0.95 \columnwidth]{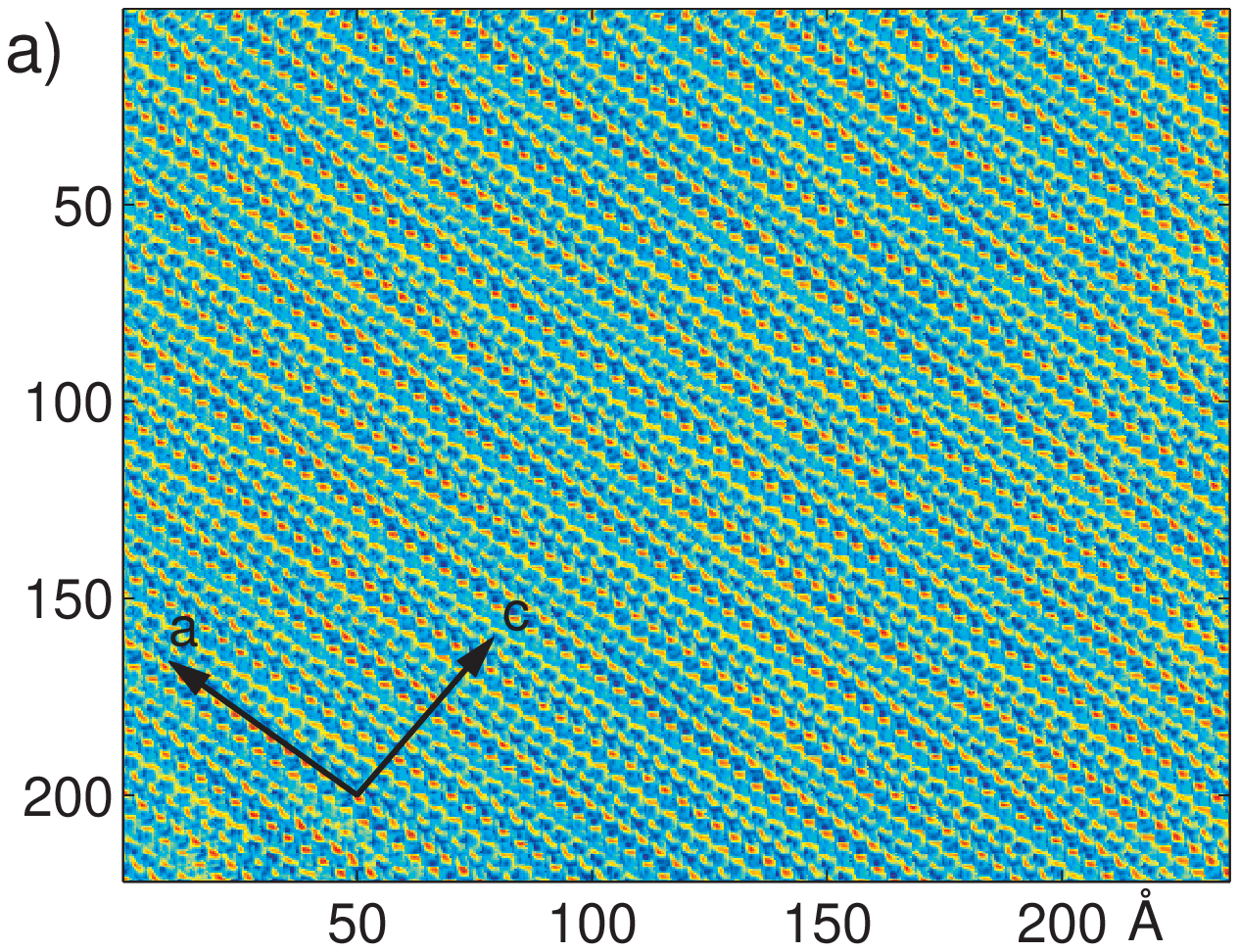}
\includegraphics[width=0.95 \columnwidth]{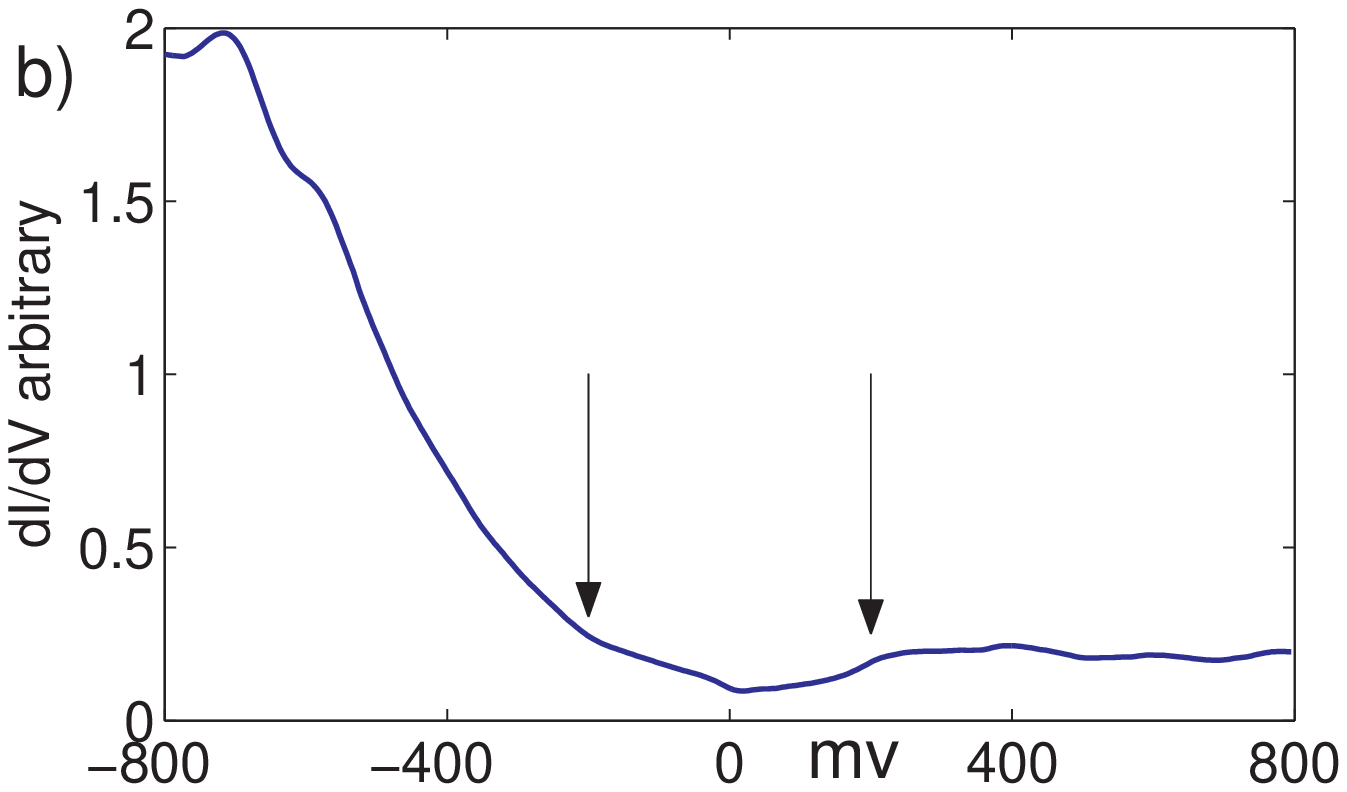}
\vspace{-4mm}
\caption{a) Topography taken with a  voltage bias of +200 mV, 50 pA current. b) Average spectrum in the range $\pm$800 mV. Arrows indicate kinks in the spectra near the gap edge.}
\label{scan}
\end{figure}

In this paper, we use scanning tunneling  microscopy and spectroscopy to study the unidirectional  CDW system TbTe$_3$.  Large area scans using a range of bias voltages allow us to resolve the ambiguity regarding the wave-vectors that comprise the CDW and show evidence of its fully incommensurate  nature.  Topographic data at different bias voltages highlight two spontaneous symmetry breaking effects in breaking the lattice point group symmetry: formation of the CDW and dimerization. 

We performed measurements on a home-made UHV cryogenic STM.  Single crystals were grown using a self-flux method as previously reported \cite{ru-transport}.  The samples were cleaved between the two Te planes in $< 2\times 10^{-10}$ Torr vacuum and quickly lowered to the $\sim$6 K section of the microscope, where cryo-pumping ensures that the surface remains free from adsorbates.  Topography were taken at several bias voltages ($=V_{sample}-V_{tip}$) and setpoint currents (typically 100pA or less).  Scan sizes were as large as $\sim 240 \times 240$ \A$^2$.  While surfaces often have large areas with no obvious surface impurities (which might pin the CDW), flakes or other "dirt" every few hundred \A~limits the maximum size of our scans. Thermal compensation, hysteresis minimization techniques, and post processing were used to reduce the amount of scan distortion in the images.   An example of such a scan with bias voltage +200 mV is shown in Fig.~\ref{scan}a. 
We also took spectroscopic scans over smaller areas, i.e.~a $dI/dV$ (proportional to the local density of states) spectrum at every point.  The spectra mainly had spatial variations relating to the lattice and CDW, thus we show in Fig.~\ref{scan}b an averaged spectrum taken in the range of $\pm$ 800 mV.  The CDW gap is marked with arrows. States in the gap are clearly seen, indicating that the FS is only partially gapped. The maximal gap was measured at $\approx$ 240 mV by photoemission. \cite{brouet-private}

\begin{figure}[h]
\includegraphics[width=1 \columnwidth]{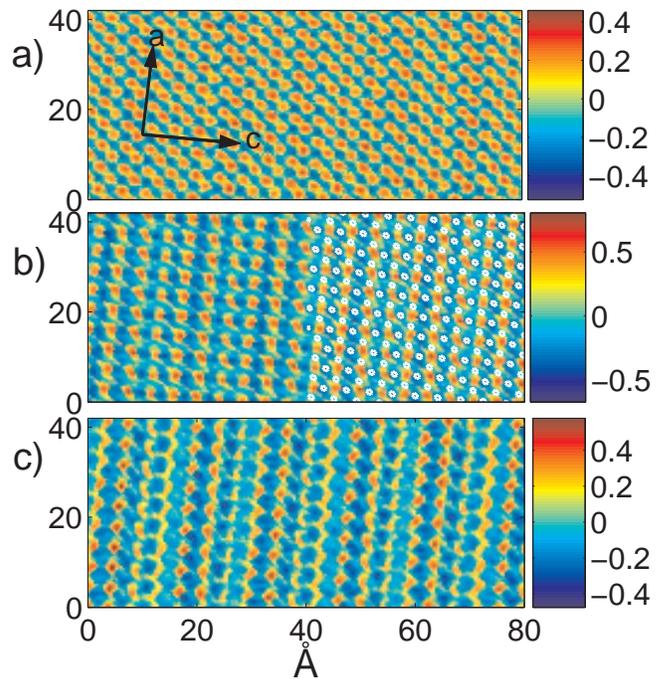}
\vspace{-4mm}
\caption{Zoomed-in view of topographic scans with 50 pA setpoint current.  All units in \AA.  a)Topography at -800mV b) -200mV. Right half shows locations of the surface Te atoms. c) +200mV.  }
\label{topo1}
\end{figure}

For STM, CDWs should appear in topography.  However, different sample bias may give  different results depending on how many of the states responsible for the CDW are integrated.  Fig.~\ref{topo1}a shows a (zoomed in) scan at -800mV and 50 pA, a voltage outside the gap which should include all the states responsible for CDW formation.   We observe the square lattice of the surface Te layer, with an observed Te-Te (average) spacing of $\approx$ 3 \A, in good agreement with bulk diffraction measurements \cite{malliakas}.  The atomic features visually swamp any CDW modulation.  One possible explanation for the relatively small amplitude of the CDW signal is that at these large negative energies, (and relatively low tunnel currents) we are probing the filled $p_z$ orbitals of the Te atoms.  Their extended nature causes the tip to be further away from where the CDW wavefunctions exist.  Since the CDW  wave function decays exponentially out of the plane  \cite{sacks}, its  contribution to the topography is expected to be small.  The second is that it is energetically unfavorable to have large charge inhomogeneities due to Coulomb repulsion, and thus when looking at quantities that are closer to representing total charge, i.e. large bias voltages, the CDW amplitude appears small.  

\begin{figure}[h]
\includegraphics[width=1\columnwidth]{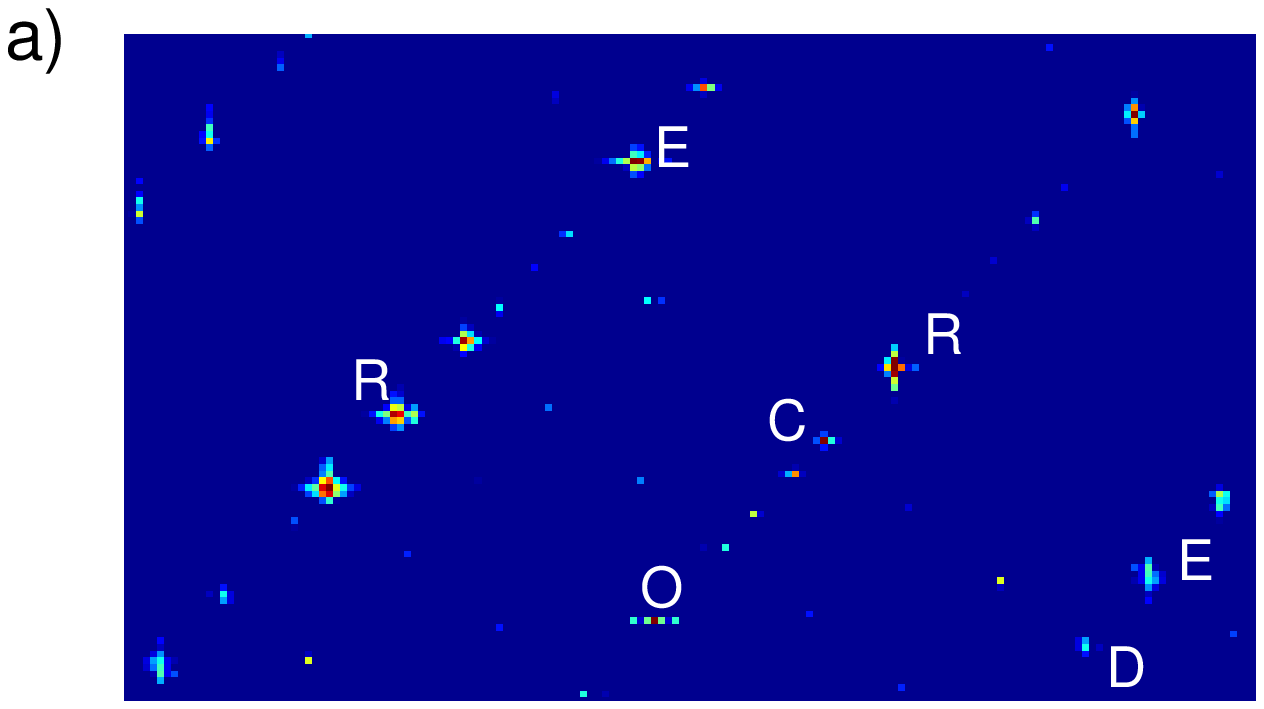} 
\includegraphics[width=.9\columnwidth]{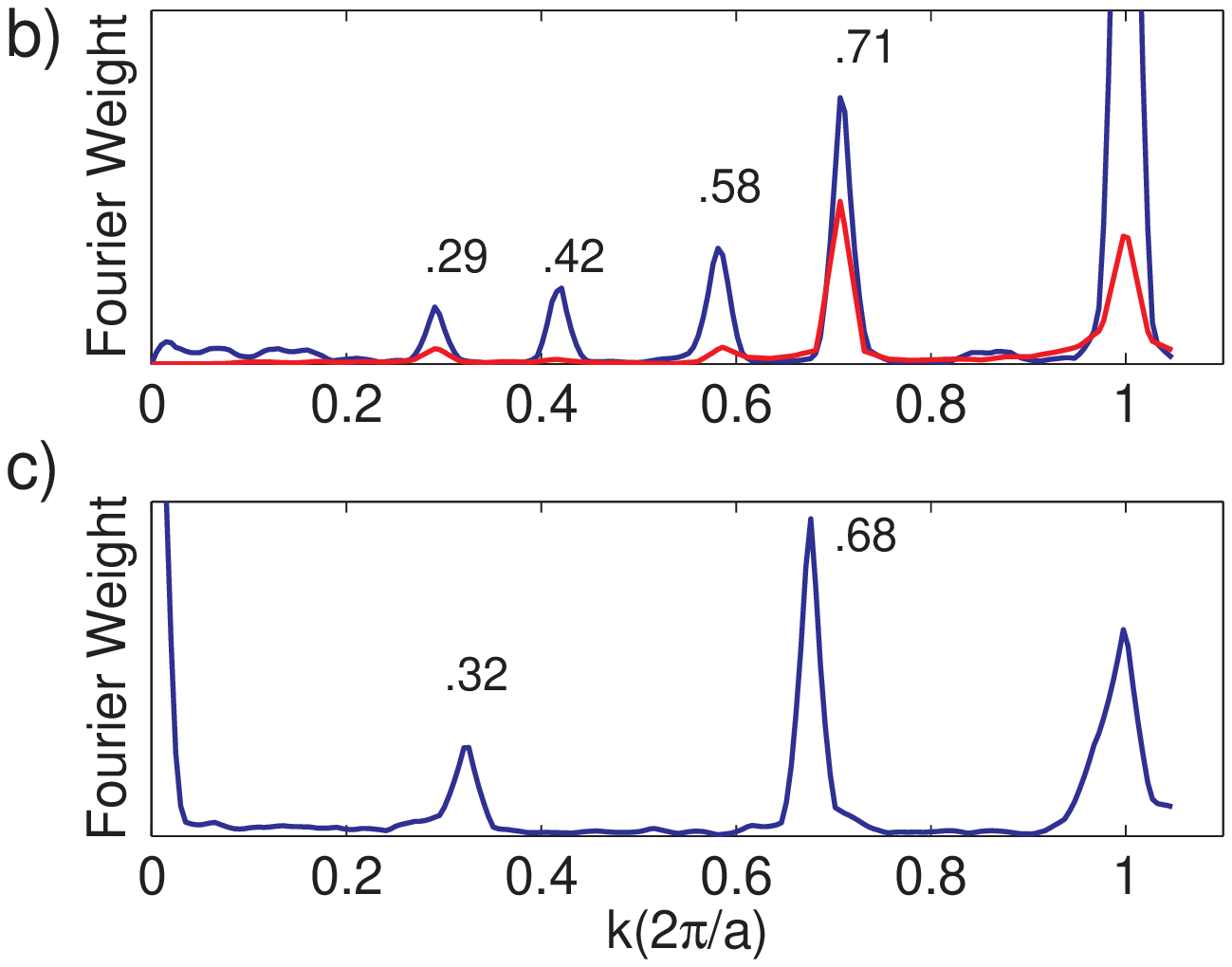} \caption{(Color) (a) FFT of +200 mV topography (Fig.~\ref{scan}(a)).  O = Origin; E = Te square lattice (2DBZ); R = block layer (3DBZ).  (b) Blue: Line cut from O to R, thru the CDW.  Red: same analysis for -800 mV topograph. Amplitude rescaled for clarity. (c) Line cut from the CDW point C to D.}
\label{fft}
\end{figure}

We also note the presence of dimerization, with pairs of atoms connected in an upper-left, lower-right direction.  First, this represents another broken symmetry, as it chooses a direction 45$^o$ (as opposed to -45$^o$) to the CDW.   Second, this is another mechanism for the unit cell doubling seen in this material. 
Although the Te net is expected to be unstable against a 3.1 \AA~  bond length \cite{patschke}, this effect has not been directly observed by XRD, but has been inferred in the PDF performed in ref \cite{michigan}.
We do not see a complex pattern of oligimers as proposed in the works of Malliakas {\it et al.} \cite{malliakas} since the simple two-atom dimers have a repeating pattern of long and short bonds within a row.
This is also suggested in Fig.~\ref{fft}b (red line) by a lack of lower frequency components in the FFT.  Although what we see may only be a surface effect, this observation should prompt a re-examination of the \RT3 crystal structure above and below $T_c$.

Fig.~\ref{topo1}b is data taken over the same area at -200mV, a value inside the maximal CDW gap.  Since the Fermi surface is only partially gapped, we expect to see states responsible for CDW formation at these lower energies.   We note again a unit cell doubling effect which shows up not as dimers, but as a new square lattice rotated 45$^o$. With respect to the surface Te plane,
three different mechanisms for unit cell doubling have been proposed in \RT3:  Dimerization, the adjacent block layer, and the stacking of the pair of Te layers deeper in the sample, which results in two crystallographically inequivalent Te atoms at the surface.  If measured with a real space probe, these mechanisms all give the same Fourier points for the superlattice, albeit with different registry to the surface Te atoms.
The peaks of the superlattice would fall between two Te atoms, four Te atoms, or on every other Te atom, respectively. 
Since we lack spatial registry between scans, we instead look at the phase of the Fourier peaks for the Te lattice combined with the spectroscopic data.   The locations of the surface Te atoms 
are depicted as white circles in the right half of Fig.~\ref{topo1}b.  Since the red spots fall between four Te atoms, this suggests that we are imaging effects from the Tb atoms in the block layer.  Seeing effects from the next layer below is not unheard of in STM, e.g. graphite \cite{graphite}.  For this set of data, the tip is ~3 \AA~ closer to the sample than it was at -800mV bias. Thus, the CDW and the unit cell doubling effects from the block layer show up more strongly. 

Fig.~\ref{topo1}c shows a zoomed-in topograph at +200 mV.  According to the spectra, the lower density of  states at positive bias suggests that the tip is even closer to the sample.  The irregular shapes of the "atoms" is likely due to the convolution of the true sample topography with tip states. This set of data shows most 
strongly the CDW ordering and effects of the block layer.

To understand the structure of the CDW, we compare the Fourier transforms of the +200mV and -800mV topographies.  
In Fig.~\ref{fft}a (+200mV) the square lattice 
from the Te atoms is clearly seen, as well as that of the block layer, which creates the unit-cell doubling effect as described earlier.
Fig.~\ref{fft}b (blue line) is a line cut showing five peaks- the
superlattice at $2\pi/c$, and four intermediary peaks at  0.29, 0.42, 0.58,
and .71 ${\pm 0.02} \times 2\pi/c$. 
In another
interpretation, Kim et al.\cite{michigan} identified these peaks as satellites from a discommensuration envelope  \cite{thompson}.  However, the -800 mV data (Fig.~\ref{fft}b red line) show that $q_{CDW}$=0.71, a value outside the 3DBZ, is the dominant contribution to the CDW, and that the other peaks are greatly reduced.  Thus 
they are not representative of an intrinsically discommensurate structure, as they do not exist under different tunneling conditions.  Rather, we identify them as 0.29=1-$q$, 0.42=2$q$-1, and 0.58=2-2$q$ which is mixing between $q$ and the superlattice.
These peaks result from {\it distortions} of STM DOS measurements due to the underlying block layer.  This distortion seems to increase when the superlattice point is stronger, as in the blue line.  
It is analogous to the effect seen in ARPES \cite{brouet} where the strength of coupling to the superlattice period determines the amount of spectral weight folded back.
In our data, the Fourier peaks are narrow, indicating a long coherence length for the modulations.  Additional width to the peaks comes from a combination of scan distortion, spectral leakage \cite{spec-leak}, and some variation in the CDW amplitude, possibly due to defects below the surface.  Overall, our interpretation that the "true" CDW has little harmonic content and a long coherence length is consistent with the X-Ray data\cite{ru-xray} which show a nearly perfectly sinusoidal modulation in the bulk.

\begin{figure}
\includegraphics[width=1\columnwidth]{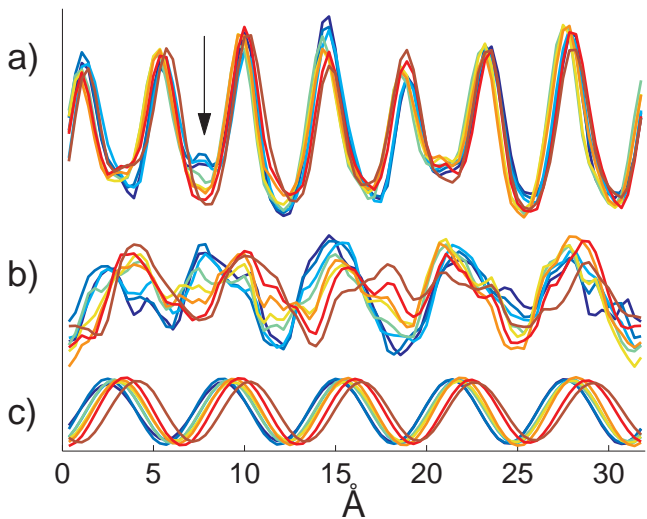} \caption{(Color) (a) Line scan along 
the CDW, in groups of 7 unit cells.  Blue is one end of the sample, evolving to dark red (other end) (b) Same signal, with atomic 
features filtered out, and amplitude multiplied by 2.  (c) Main CDW component only.} 
 \label{supercells} 
\end{figure}

Having ruled out a nearly commensurate (NC) structure, next we check if the CDW is fully incommensurate or commensurate against the nearest fraction of 5/7 (=0.714). (Although it is difficult to distinguish incommensurate against a high denominator fraction commensurate.)  Recent X-ray data shows 
  that the CDW is incommensurate in the bulk at room temperature with $q$=0.704 ${\pm 0.001}$\cite{ru-xray}.
However, it is well known that incommensurate CDW materials can change their wave-vectors as a function of temperature 
\cite{malliakas_2006} or even undergo a change to the NC or commensurate structure \cite{thompson}.  There may also be 
surface effects which make the CDW different at the surface than the bulk.  \cite{kivelson-extraordinary} 

We check our low temperature data for commensuration against 5/7 by filtering out features perpendicular to the CDW 
wave-vector, then taking a line cut in the direction of the CDW wave-vector in real space.  We then match up groups of 7 
unit cells.  If the CDW were commensurate, all the groups would be identical.  Fig.~\ref{supercells}a shows the results of 
this analysis, with the arrow pointing at a region of smooth evolution from one group to the next.  From this, we 
conclude that the CDW is incommensurate.  Fig.~\ref{supercells}b is the same data, with the lattice filtered 
out.  A $\sim$5/7 modulation can clearly be seen, with some distortion caused by the lattice positions.  This distortion creates the 
other peaks in Fig.~\ref{fft}b (blue line).  Fig.~\ref{supercells}c shows the Fourier filtered component for the CDW 
slowly advancing as expected for slight incommensuration with $q <$ 5/7. \cite{peakspacing} This value of $q$ means that the four peaks in Fig.~\ref{fft}(b) (blue line) are not evenly spaced, but rather the two center peaks are slightly spread apart.  
 Likewise, in the work of Kim et al,\cite{michigan} the two center peaks are slightly closer together, consistent with  $q > $5/7 in CeTe$_3$.  

It is not unexpected that this material would be incommensurate 
for the following two reasons: 1) The CDW is FS nesting driven, and thus does not have to be related to the lattice by a rational fraction.  2) The high order denominator ($\geq$ 7) for the nearest commensuration fraction implies that any lattice 
locking or commensuration effects are weak.\cite{gruner} In addition, resistivity and heat capacity measurements do not 
indicate any transitions from room temperature down to 1.8 K (aside from a Ne\'el transition at 5.6 K).  \cite{ru-transport} Thus we conclude that 
for R=Tb, the CDW stays incommensurate with $q <$ 5/7 down to low temperatures.  

Finally, we note weak Fourier signals at ${q_c}\approx$(0.29 or 0.71) and ${q_a}\approx$(0.32 or 0.68), as satellites perpendicular to the CDW.  Thus, 
in Fig.~\ref{fft}c we take a like cut in the direction from C to D.  This means 
that there is a $\approx 3 \times a$ modulation perpendicular to the CDW, 
and that each row of the CDW is shifted laterally with respect to the next.  
This signal is not seen in room temperature X-ray diffraction, meaning that it is 
either a previously unobserved second CDW phase transition below room temperature or a surface reconstruction.

In conclusion, we show that TbTe$_3$ has a fully incommensurate, unidirectional CDW at low temperatures, with a dominant wave vector $q$=0.71 ${\pm 
0.02}$ outside the 3DBZ.  The adjacent R-Te block layer strongly affects the topography at lower bias voltages by acting as an additional periodicity which distorts STM measurements of the CDW. Our observation of simple two-atom dimerization in the Te net is consistent with the bimodal bond length distribution suggested in powder PDF, and our interpretation of the CDW as fully incommensurate with long coherence length agrees with X-Ray data. 

 \bigskip

\noindent {\bf Acknowledgments:} We thank S. Kivelson, H. Yao and E.A. Kim  for useful discussions.   STM work supported by the U. S. DoE under contract No. DE-FG03-01ER45925.  The crystal growth was
supported by the U. S. DoE under contract No. DE-AC02-76SF00515.


\begin{references}
\bibitem{unidirectional}
We refer to the CDW as unidirectional instead of 1-D to differentiate it from CDWs that exist in linear chain compounds.
\bibitem{spec-leak}
The Fourier weight of a signal with a single frequency is spread out when a non-integer number of periods resides within the scan range.  Care must be taken when Fourier filtering and then reverse transforming, as done in Fig.~\ref{supercells}.
\bibitem{peakspacing}We can do a similar analysis to show that $q >$ 7/10.  Thus 0.700 $<q <$ 0.714.
\end{references}
\end{document}